\documentclass[letterpaper,10 pt, conference]{IEEEtran}
\usepackage{amsmath,amssymb}
\usepackage{array} 
\usepackage[dvips]{graphicx}
\usepackage{xspace}
\usepackage{lscape}
\usepackage{here}
\usepackage{epsfig}
\usepackage{multirow}
\usepackage{color}
\usepackage{epstopdf}

\newcommand{\beq}{\begin{equation}}
\newcommand{\eeq}{\end{equation}}
\newcommand{\mx}[2]{\left[\begin{array}{#1}#2\end{array}\right]}
\newcommand{\barr}[2]{\begin{array}{#1}#2\end{array}}
\newcommand{\ds}{\displaystyle}
 \newcommand{\R}{\rm{I\kern-2pt R}}

\newtheorem{thrm}{\bf Theorem}
\newtheorem{rmrk}{{\bf Remark}}
\newtheorem{ex}{\bf Example}
\newtheorem{crllry}{{\bf Corollary}}
\newtheorem{defi}{\bf Definition}
\newtheorem{prpstn}{{\bf Proposition}}
\newtheorem{lemme}{{\bf Lemma}}

\newenvironment{remark}{\begin{rmrk} \rm}{\end{rmrk}}

\newenvironment{definition}{\begin{defi} \rm}{\end{defi}}
\newenvironment{proposition}{\begin{prpstn} \rm}{\end{prpstn}}

\IEEEoverridecommandlockouts 

\begin{document}
\title{\vspace{4mm} Multi-agent systems with CBF-based controllers -- collision avoidance and liveness from instability*} 

\author{ Mrdjan Jankovic, Mario Santillo, Yan Wang
\thanks{ *A portion of the material in this paper has been presented at the 2021 Conference on Decision and Control \cite{liveness}.

M. Jankovic, M. Santillo, and Y. Wang are with
  Ford Research and Advanced Engineering,
  2101 Village Road, Dearborn, MI 48124, USA,   {\tt\small e-mail: mjankov1@ford.com, msantil3@ford.com, ywang21@ford.com}}%
}

\maketitle

\begin{abstract}
\noindent 
Assuring system stability is typically a major control design objective. In this paper, we present a system where instability provides a crucial benefit. We consider multi-agent collision avoidance using Control Barrier Functions (CBF) and study trade-offs between safety and liveness -- the ability to reach a destination without large detours or gridlock. We compare two standard decentralized policies, with only the local (host) control available, to co-optimization policies (PCCA and CCS) where everyone's (virtual) control action is available. The co-optimization policies compute control for everyone even though they lack information about others' intentions.  For comparison, we use a Centralized, full information policy as the benchmark.  One contribution of this paper is proving feasibility for the Centralized, PCCA, and CCS policies. Monte Carlo simulations show that decentralized, host-only control policies and CCS lack liveness while the PCCA policy performs as well as the Centralized. Next, we explain the observed results by considering two agents negotiating the passing order through an intersection. We show that the structure and stability of the resulting equilibria correlates with the observed propensity to gridlock -- the policies with unstable equilibria avoid gridlocks while those with stable ones do not. 
\end{abstract}

\section{Introduction}
\noindent
With the recent advances in automated driver assist systems, autonomous vehicles, and multi-robotic systems, management of agent-to-agent interactions has received a lot of attention. Each agent must be capable of planning and executing paths in real time while assuring collision-free operation. A challenging situation could arise from mixed operating scenarios such as with multi-brand, multi-robot factories or heterogeneous driving scenarios with fully autonomous, semi autonomous, and human-driven agents/vehicles that both compete and cooperate. This can lead to complex feedback loops that are only partially controllable from each agent's perspective, presenting an opportunity for the rigors of feedback control. 

In recent years, Control Barrier Functions (CBF) \cite{ames_pp,ames_cbf,wieland} have shown great promise in providing a computationally efficient method that is both provably safe and able to handle complex scenarios. Similar to Model-Predictive Control (MPC), CBF is a model-based control design method that can be formulated as a quadratic program (QP) and solved online using real-time capable solvers, e.g. \cite{odys}. A few noteworthy differences between CBF and MPC include (i) MPC generally relies on a set of linearized dynamic systems to cover the nonlinear model range, whereas CBF deals with the nonlinear (but control affine) model directly, (ii) for non-convex constraints, MPC requires convexification, sequential convex programming, or mixed integer programming, while CBFs do not see them as such,  and (iii) MPC takes advantage of future state prediction whereas CBF does not.

In a controlled environment with all agents having the ability to communicate, a centralized controller -- an off-board computer that takes in all agents' inputs (such as their desired acceleration), calculates and relays the optimal action for each agent -- may be employed \cite{wilson}. One contribution of this paper is a proof that the centralized, distance-CBF based QP problem is always feasible.

In less controlled scenarios such as vehicles operating on roadways, or multi-brand robots operating without common communication protocols, a decentralized controller may be necessary. In this case, each agent computes and executes the best control for itself given the information available similar to how we drive vehicles. Several variations of the decentralized CBF policy exist. In particular, the Decentralized Follower  (DF) method \cite{borrmann} assumes each agent takes full responsibility for collision avoidance while the Decentralized Reciprocal (DR) method \cite{wang} assigns each agent a fraction of responsibility. Robust CBF (RCBF) \cite{jankovic_aut} was used as the basis for the development of the Predictor-Corrector for Collision Avoidance (PCCA) algorithm \cite{arXiv}. We herein present a novel decentralized controller, called Complete Control Set (CCS), that assigns appropriate agent responsibility and guarantees constraint adherence in the two-agent case. The PCCA and CCS are performing ``co-optimization" -- they compute the best course of action for every agent with local, incomplete information. The advantage is that the corresponding QPs are always feasible. The difference between them is that CCS simply discards local copies of other agent actions, while PCCA compares them to the observed actions -- hence the "predictor-corrector" nomenclature -- and feeds the difference back as a disturbance. 

After introducing each algorithm, we assess  feasibility and its effect on online algorithm implementation. We then proceed to compare algorithm metrics on liveness, collisions, and feasibility through a randomized five-agent Monte-Carlo simulation trials for each method (each set up identically). It turns out that the two decentralized algorithms and the CCS exhibit a certain percent of gridlocks as well as generally slower arrival to the destination compared to the Centralized and PCCA policies which showed consistently fast convergence with no gridlocks. 

As  mentioned above, agents operating independently in a shared space create complex feedback loops. We conjectured that 
the equilibrium structures and their stability play a decisive role in determining propensity to gridlock.  Gridlocks 
for multi-agent systems have been studied before in robotics literature. The results typically
deal with decentralized policies (DF, DR) and propose methods to deconflict the agents once a gridlock is detected -- see, for example
\cite{ celi, duhaut, grover, wang} and references therein.  Here, we are interested in trying to explain the observed differences between
the control policies. We analyze the equilibrium structure in the joint space for a simple problem of two agents negotiating 
 passing order through an intersection or a merge point. The decentralized policies as well as CCS have
a stable set of equilibria which explains their propensity to gridlock. In contrast, the Centralized and
PCCA policies have unstable equilibria and thus only a set of measure zero that ends up in a gridlock. 
Moreover, the instability is exponential, resulting in fast movement away from the gridlock point even for trajectories that start
close to the stable manifold of the equilibrium set. 

The paper is a significant extension of the conference paper \cite{liveness}. Besides many improvements and clarifications throughout the text and added
figures better explaining Monte Carlo results in Section \ref{sec:sims}, the main addition is the equilibrium analysis in Section \ref{sec:instability}. This section contains an explanation of the observed gridlock propensity linking it with policy dependent stability of multi-agent systems. 

The rest of this paper is organized as follows. Section \ref{sec:rcbf} reviews CBF and RCBF based control. Section \ref{sec:agents} introduces the dynamic model for the agents. Section \ref{sec:algs} reviews or introduces CBF-based controllers for collision avoidance. The simulations in Section \ref{sec:sims} consider randomized trials of five interacting agents with a stationary obstacle. The equilibrium analysis is provided in Section \ref{sec:instability}.  \\

\noindent
{\bf Notation:} For a differentiable function $h(x)$ and a vector $f(x)$, 
$L_fh(x)$ denotes $\frac{\partial h}{\partial x} f(x)$.
A continuous function $\alpha(\cdot)$ is of class $\cal{K}$ if it is strictly increasing and satisfies $\alpha(0)=0$. We additionally assume $\alpha \in \mathcal K$ is Lipschitz continuous. A function $\gamma(t,\varepsilon)$ is said to be ${\mathcal O}(\varepsilon)$ if $|\gamma(t,\varepsilon)| \le \kappa |\varepsilon|$ for some $\kappa>0$  and for all sufficiently small $\varepsilon$.

\section{Robust Control Barrier Functions reviewed}\label{sec:rcbf}
\noindent
In this section, we briefly review the concepts of CBFs -- introduced in \cite{wieland} and later combined with quadratic programs (see, e.g. \cite{ames_pp,ames_cbf}) -- and of RCBFs introduced in \cite{jankovic_aut}. CBFs apply to nonlinear systems affine in the control input
\beq  \dot x = f(x) +g(x) u \label{nls} \eeq
with $x\in {\R}^n$, $u \in {\R}^m$, $f(x)$ and $g(x)$ Lipschitz continuous. RCBFs extend CBFs to systems with a bounded external disturbance $w(t) \in {\R^\nu}$, $ \| w(t)\| \le \bar w > 0$, of the form 
\beq \dot x= f(x) + g(x) u + p(x) w \label{dyn_w_dist} \eeq
where $p(x)$ is also Lipschitz continuous.

One control objective is to regulate the system
 to the origin or suppress the disturbance (i.e. achieve input-to-state stability (ISS)) 
and we assume that there is a known baseline controller 
$u_0$ that achieves the objective. The other control objective is to
keep the state of the system in an admissible (i.e. safe) set defined by ${\mathcal C} = \{x \in {\R}^n: h(x)  \ge 0\}$
where $h(x)$ is a twice continuously differentiable function. 
Here we combine definitions of CBF and RCBF into one. \\

\begin{definition} ({\em  CBF and Robust-CBF}) \
A twice continuously differentiable function $h(x)$ is a CBF for the system (\ref{nls}) 
if there exists a function $\alpha_h \in \mathcal K$ such that 
\beq L_gh(x) = 0 \ \Rightarrow  \ L_fh(x) + \alpha_h(h(x)) > 0 \label{cbf} \eeq
The function $h(x)$ is an 
RCBF for the system (\ref{dyn_w_dist}) if
\beq L_gh(x) = 0 \ \Rightarrow  \ L_fh(x) - \|L_p h\| \bar w + \alpha_h(h(x)) > 0 \label{r-cbf} \eeq
\end{definition}

In the CBF case, the definition asks that, when the control over the evolution of $\dot h = L_fh +L_gh u $ is lost ($L_gh = 0$),
the rate of decrease of $h$ to 0 is not faster than $\alpha(h)$. Similarly, for the system with disturbance, the bound on the 
rate of decrease applies for the worst case disturbance $ \bar w$. 

One advantage of CBFs for control affine systems is that they naturally lead to linear constraints on the control input $u$ 
that could be enforced online. A quadratic program (QP) is set up to enforce
the constraint, while staying as close as possible to the baseline (performance) control input $u_0$:  \\

\noindent
{\bf (R)CBF QP Problem}: Find the control $u$ 
 that satisfies
\beq \barr{l}{\ds \min_u \|u-u_0\|^2 \ \  {\rm subject \  to}  \\*[2mm]
 \ds F_i \ge 0, \ \  i = 0, 1,\  {\rm or}\  2 }\label{rQP} \eeq
where we select $F_0 = L_fh(x) + L_g h(x) u + \alpha_h(h(x)) $ if $h$ is a CBF for the system (\ref{nls});
$F_1 = L_fh(x) - \|L_p h(x)\| \bar w + L_g h(x) u + \alpha_h(h(x))$ if $h$ is an RCBF for the system (\ref{dyn_w_dist}) with an unknown disturbance; or $F_2=
L_fh(x) + L_p h(x) \hat w + L_g h(x) u + \alpha_h(h(x))$  when an estimate/measurement
of the disturbance is available. \\

The available results (e.g. \cite{ames_pp, ames_cbf, jankovic_aut}) guarantee that the resulting control is Lipschitz continuous, the barrier constraint $F_i$ 
is satisfied, which implies that $h(x(t)) \ge 0, \forall t$ and the safe set $\mathcal C$ is forward invariant. Note that strict ``$>$'' is needed in the (R)CBF definition (\ref{cbf}) and (\ref{r-cbf}) to guarantee
Lipschitz continuity of the control law \cite{jankovic_aut}, or,  alternatively, $L_gh(x) \not = 0, \forall x \in {\mathcal C}$ needs to be assumed as in \cite{ames_pp}. 

In the distance-based barrier functions considered in the rest of the paper, the control input does not 
appear in their first derivative as assumed in the definition of (R)CBF. The inputs appear in the second derivative of $h$, so we follow the 
ideas of \cite{nguyen, xu_rel_deg} for dealing with higher relative degree:  
instead of enforcing $\dot h + \alpha_h(h) \ge0$, we switch to linear barrier dynamics and enforce
\beq \ddot h + l_1 \dot h + l_0 h\ge 0 \label{r2_barrier}\eeq
as the QP constraint. The parameters
$l_0, l_1$ should be selected so that the two roots $\{-\lambda_1, -\lambda_2\}$   of the polynomial
$s^2+l_1 s + l_0 = 0$ are negative real ($\lambda_{1/2} = \frac{l_1\pm\sqrt{l_1^2 -4l_0}}{2}$ ). 
It is a matter of straight-forward calculation to show that, if the barrier constraint (\ref{r2_barrier}) holds,  the set 
${\mathcal C}^* = \{ (x): h(x) \ge 0, h(x) \ge -\frac{1}{\lambda_l}\dot h(x) \}$, where $-\lambda_l$ is either of the two eigenvalues,
is forward invariant. With ${\mathcal C}^*\subset {\mathcal C}$ the original constraint $h(x) \ge 0$ will be satisfied and $\mathcal{C}^* \rightarrow \mathcal{C}$ as $\lambda_l \rightarrow \infty$.
 
For the second-order barrier, the QP constraints that need to be enforced for ${\mathcal C^*}$ to be forward invariant are 
\beq  F_0 = L_f^2 h + L_gL_f h u + l_1 L_fh + l_0 h \ge 0 \label{r2-constr} \eeq
in the case of the CBF for the system without disturbance;
\beq F_1 = L_f^2 h - \|L_pL_fh \|\bar w + L_gL_f h u + l_1 L_fh + l_0 h \ge 0 \label{r2_const_barw} \eeq
for an RCBF with unknown disturbance bounded by $\bar w$;  or
\beq F_2 = L_f^2 h + L_fL_p h \hat w + L_gL_f h u + l_1 L_fh + l_0 h \ge 0 \label{r2_const_hatw} \eeq
for an RCBF with known disturbance estimate $\hat w$.

\section{Holonomic Agent Model}\label{sec:agents}
\noindent
In the literature, agents are typically modeled either with holonomic double integrators in the X-Y plane or as a non-holonomic ``unicycle'' or ``bicycle'' model. For simplicity, here we consider the first option.
An agent $i$ is modeled as a disk of radius $r_0$ with the center motion
given by  the double integrator in each dimension:
\beq \barr{l} {\dot x_i= v_{xi} \\
         	     \dot y_i = v_{yi} \\
	     	      \dot v_{xi} = u_{xi}, \\
		      \dot v_{yi} = u_{yi}} \label{agent_i} \eeq

The relative motion between any two agents $i$ and $j$ is given by
\beq \barr{l}{ \dot \xi_{ij}= v_{ij} \\ 
\dot v_{ij} = u_i - u_j} \label{diff_motion} \eeq
where $\xi_{ij} = [x_i - x_j, y_i - y_j]^T$ is the center-to-center (vector)
displacement between the two agents, $v_{ij} = [v_{xi} - v_{xj}, v_{yi} - v_{yj}]^T$ 
is their relative velocity, and $u_i = [u_{xi}, u_{yi}]^T$ is agent $i$'s control.
Our goal is to keep the $\|\xi_{ij}\|$ larger than $r \ge 2r_0$ (with the  distance $r$ strictly greater than $2 r_0$ we are providing a ``radius margin" relying on
robustness of barrier functions (see \cite{xu}) to push the states out of the inadmissible set).  
To this end, we define a barrier function
\beq h(\xi_{ij}) = \xi_{ij}^T\xi_{ij} - r^2  \label{h_ij} \eeq
with the goal to keep  it greater than 0. The advantages of this barrier function over alternative ones used for multi-agent collision avoidance 
is that (i) we can prove the feasibility of the centralized and a few decentralized QPs and (ii) it allows
a radius (barrier) margin because the calculation does not collapse when $h(\xi_{ij} )<0$. 
A disadvantage is that $h$ has relative degree two from all 4 inputs. Because of this, 
we apply the approach described in Section \ref{sec:rcbf} and form a CBF barrier constraint:
\beq  F_{ij} := \ddot h + l_1 \dot h + l_0 h = a_{ij} + b_{ij}(u_i - u _j )\ge 0  \label{h_constr1} \eeq
where $a_{ij} = 2v_{ij}^Tv_{ij}+ 2 l_1 \xi_{ij}^T v_{ij} +  l_0 (\xi_{ij} ^T\xi_{ij} - r^2)$, $b_{ij}  = 2 \xi_{ij} ^T$, 
and $u_i$ and $u_j$ are the control actions of the two
agents. The function $h$ is a CBF for the system (\ref{diff_motion}) because $L_gh= 2b_{ij} \not = 0$ unless the two agents completely overlap
(that is, are well past the point of collision). As a result, we can always enforce positive invariance of the 
admissible set ${\mathcal C}^*_{ij}= \{ (\xi_{ij}, v_{ij}): h(\xi_{ij}) \ge 0, h(\xi_{ij}) \ge -\frac{1}{\lambda_l }\dot h(\xi_{ij}, v_{ij}) \}$ , with $\lambda_l$ one of the two eigenvalules, as discussed above. Without loss of generality, we will use $l = 1$.

As we shall see below, with the centralized controller we can guarantee that the agents do not collide even if there is no barrier margin:
$r = 2r_0$. In the non-ideal case -- for example, discrete-time implementation, the only way a QP could be implemented --
some very small barrier function violations appear. Non-centralized controllers might have larger violations because
two agents $i$ and $j$ compute $u_i$ and $u_j$ independently, based on different information available to them. 
To the extent they don't agree, a difference between the left hand sides of the centralized barrier constraint $F_{ij}^c \ge 0$, which guarantees collision avoidance, 
 and the actual $F_{ij}^a \ge 0$ resulting from independent control calculations could appear.  On the other hand, the radius margin creates a repelling term 
 equal to $2 l_0(r^2 - r_0^2)$ on the boundary of the actual barrier function we want to enforce, $h_0= \|\xi_{ij}\|^2 -2r_0$, providing a degree of robustness (see \cite{xu} for a more general consideration).

\section{CBF-Based Collision Avoidance Algorithms}\label{sec:algs}
\noindent
In the absence of other agents, we assume that each agent has its own preferred control action $u_{0i}$ (for agent $i$), 
computed independently of the collision avoidance algorithm. In this paper,  each agent knows its own final destination 
and uses the Linear Quadratic Regulator (LQR) controller to compute $u_{0i}$ that is supposed to bring it there. On the other hand,
we set up the Centralized controller that controls all the agents, knows everyone's $u_{0i}$'s, 
and uses the barrier constraints between each two agents defined in the previous section.
Thus, the Centralized controller could be set up as the solution to the following quadratic program:\\

\noindent
{\bf Centralized QP}: Find the controls $u_i, i=1,\ldots, N_a$ 
\beq \barr{l}{\ds \min_{u_{1}, \ldots u_{N_a} } \sum_{i=1}^{N_a} \|u_i -u_{0i}\|^2 \ \  {\rm subject \  to}  \\*[2mm]
 \ds a_{ij} + b_{ij}(u_i - u_j )\ge 0  \ \ \forall i,j =1,\ldots, N_a , i\not = j}\label{rQPc} \eeq
where $a_{ij}$ and $b_{ij}$ are defined in the previous section (after equation (\ref{h_constr1})) and $N_a$ is the number of agents. 

The QP solution could be computed by a central node and communicated to the agents, or each agent could solve 
the QP independently, which still requires communication between them because they need to know each-others' baseline controls $u_{0j}$. If this QP problem is feasible, and this is proven below, the control action would satisfy all the  barrier constraints (\ref{h_constr1}) and guarantee collision-free operation (see  \cite{wang}).

Without communication, the base control action $u_{0j}$ for the target (i.e. other) agents are not available to the host  $i$ (the agent doing the
computation). In this case, each agent could implement an on-board decentralized controller.  One version, included here because it
resembles many defensive driving policies considered in  literature, is for each agent to accept full responsibility for 
avoiding all the other agents. Borrowing nomenclature from  
Game Theory, we refer to this policy as ``Decentralized Follower'' (DF): \\

\begin{figure}[t!]\vspace{-.05in}
    \centering
\includegraphics[scale=0.40]{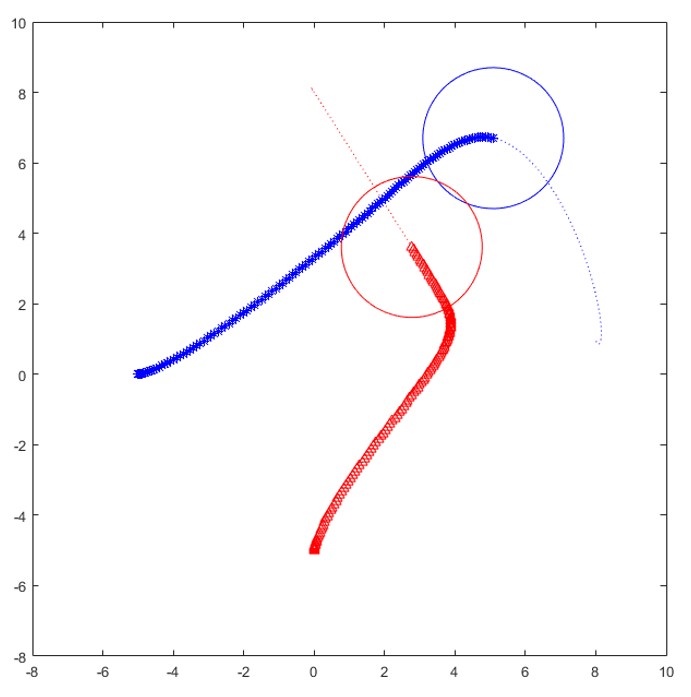}
    \vspace{-0.1in}
    \caption{Two agents colliding while crossing paths with the Decentralized Follower policy.}
    \label{fig:DFA}
    \vspace{-0.15in}
\end{figure}

\noindent
{\bf Decentralized Follower QP} (for agent $i$): Find the control $u_i$ for the agent $i$
 that satisfies
\beq \barr{l}{\ds \min_{u_{i}}  \|u_i -u_{0i}\|^2 \ \  {\rm subject \  to}  \\*[2mm]
 \ds a_{ij} + b_{ij}u_i\ge 0  \ \forall  j =1,\ldots, N_a , j\not = i}\label{rQPd} \eeq 
This formulation is essentially the same as in \cite{borrmann} but a different barrier function is used, as described above.
The agent $i$ has only its own actions ($x$ and $y$ accelerations) to avoid all other agents and 
there are no guarantees that the DF QP is feasible. Even when it is feasible and the agents apply the same defensive algorithm, 
there are no collision avoidance guarantees. The reason is that each agent knows only its own acceleration $u_{0i}$, 
and may assess it safe to apply. That is, if $a_{ij}+b_{ij}u_{0i} \ge 0$, $\forall j$, agent $i$ would consider $u_{0i}$ safe to apply.
Similarly, agent $j$ might find that $u_{0j}$ is safe to apply. However, $a_{ij}+b_{ij}u_{0i} \ge 0$
and $a_{ij} - b_{ij}u_{0j} \ge 0$ does not imply $a_{ij}+b_{ij} (u_{0i} - u_{0j}) \ge 0$, that, if satisfied, would actually guarantee
collision avoidance. Indeed, our simulations show that, even with only two agents, there could be a collision as seen 
by overlapping circles in Figure \ref{fig:DFA}. In the simulation run shown, each agent implemented the (identically tuned) navigation policy described by (\ref{rQPd}) with their $u_{0i}$'s coming from an LQR controller.
%

We note that (i) the collision did not happen during braking (i.e. when the agents are
approaching one another), but when they both optimistically assume it is safe to accelerate; (ii) even though the algorithm 
updates controls every 50ms with new position and velocity information, the collision is not avoided when 
the actual constraint $a_{ij}+b_{ij} (u_{i} - u_{j}) \ge 0$ is violated. 

To  improve the DF performance, the  ``Decentralized Reciprocal" (DR) policy was introduced in \cite{wang}:\\

\begin{figure}[t!]\vspace{-.1in}
    \centering
\includegraphics[scale=0.37]{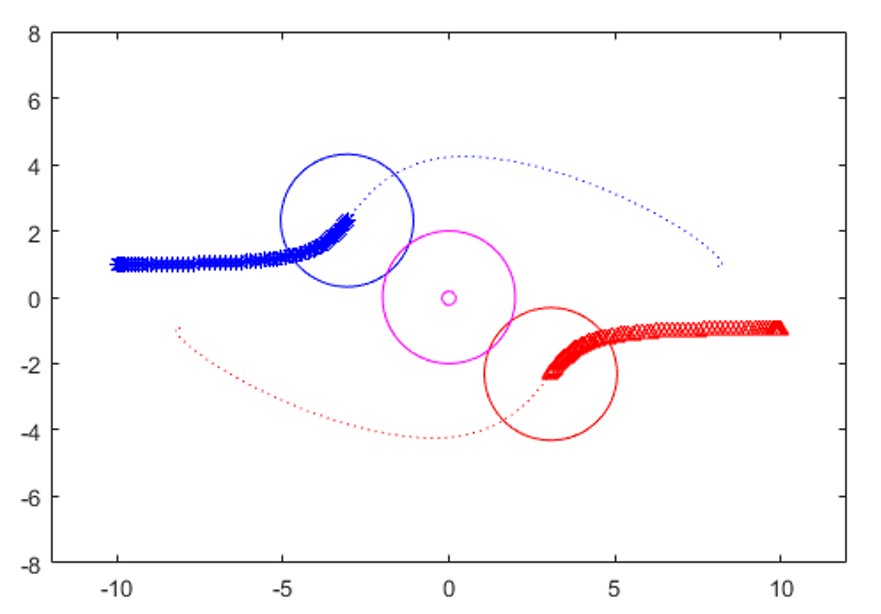}
    \vspace{-0.1in}
    \caption{Two agents passing a stationary one from opposite directions, all running the Decentralized Reciprocal policy.}
    \label{fig:DRA}
    \vspace{-0.1in}
\end{figure}

\noindent
{\bf Decentralized Reciprocal QP} (for agent $i$): Find the control $u_i$ for the agent $i$
 that satisfies
\beq \barr{l}{\ds \min_{u_{i}}  \|u_i -u_{0i}\|^2 \ \  {\rm subject \  to}  \\*[2mm]
 \frac{1}{2} a_{ij} + b_{ij}u_i\ge 0  \ \  \forall  j =1,\ldots, N_a , j\not = i}\label{rQPr} \eeq 
 The only difference from the DF version is the $\frac{1}{2}$ factor multiplying $a_{ij}$, meaning that
 each agent assumes half the responsibility for avoiding the collision (we assumed all the agents are the same). 
 The method was shown in \cite{wang}
 to guarantee constraint adherence and, hence,  collision avoidance as long as it is feasible and, when it is not, 
proposed a braking action. Braking, however, works only if all agents, even those with 
 feasible QP, apply it at the same time. To illustrate the issue, we consider two moving agents passing the 
 stationary one in the middle as shown in Figure \ref{fig:DRA}. The DR QP turns out infeasible for the agent in the middle, 
 but, because it was already stationary, the braking applied has no effect. The other two have feasible QPs and keep applying the solutions.
 This leads to collisions as shown in Figure \ref{fig:DRA} because the expected
 half contribution towards avoiding collisions by the agent in the middle has not been met.
 One possible approach to avoid the problem is for each agent to run a separate DR-QP's for itself and all the other agents and
 brake when any one of the QPs turns infeasible (no need to know $u_{0j}$'s to assess feasibility). This would increase 
 computational footprint, but also  raise the issue of when to stop braking: as soon as all QP's become feasible, or
 only after all agents have stopped. The former might lead to a jerky motion, while the latter would suffer from reduced liveness. 
 In the simulation section, we have allowed the QP solver to resolve the feasibility issue by selecting control with the 
 smallest constraint violation counting on eventual application of the radius margin.

Instead of $N_a$ DR-QPs being solved by the host to assess feasibility for all the other agents, one could consider setting up a
single QP that includes all the constraints. The problem, of course,  is that the host does not know other agents' preferred accelerations, so 
zeros are used instead of unknown $u_{0j}$'s. We refer to 
this policy as the ``Complete Control Set" (CCS): \\

\noindent
{\bf CCS QP} (for agent $i$): Find control actions $u_{ij}, j=1,\ldots, N_a$ 
such that
\beq \barr{l}{\ds \min_{u_{i1}, \ldots, u_{iN_a}}  \sum_{j=1}^{N_a} \|u_{ij}\|^2 \  {\rm subject \  to}  \\*[2mm]
 \ds a_{ij} + \rho_i b_{ij} u_{0i} + b_{ij}(u_{ii}- u_{ij})\ge 0  \ \forall  j =1,\ldots, N_a , j\not = i \\*[2mm]
 \ds a_{jk} + b_{jk}(u_{ij} - u_{ik} )\ge 0  \ \ \forall  j,k =1,\ldots, N_a \\*[2mm]
 \ds \hspace*{2.0in} j\not = k \ {\rm and} \ j,k \not = i}\label{rQPccs} \eeq
where $\rho_i$ is a design parameter used to assure that when the constraint is active for multiple agents, the actual barrier constraint (\ref{h_constr1}) is satisfied. The agent $i$ uses the CCS policy to find the control action for all the agents and implements its own: $u_i = u_{ii}^* + u_{0i} $ where
$u_{ii}^*$ denotes the solution to the QP problem.
The CCS QP is guaranteed to be feasible
(see Proposition \ref{prop_feasibility} below).

To better understand the $\rho_i$ term in front of $b_{ij} u_{0i}$, first note that $\rho_i = 1$ would be 
equivalent to the centralized QP (\ref{rQPc}) with all $u_{0j}$, except $j = i$, (i.e. all the unknown ones) set to 0
 and with a variable change for $u_{ii}$. While the complete set of
 constraints guarantees that agents will correctly split the responsibility for $a_{ij}$, other agents do not know about $u_{0i}$ so 
 the responsibility could not be split. The multiplier $\rho_i = 2$ works in the case of two agents implementing the CCS policy, when 
 constraint adherence could be established using the closed-form solution from \cite{arXiv}.
In the multi-agent case, this is not the case -- one could find situations when the constraints will not be satisfied. Still, because it is always feasible and,
as it turned out, somewhat more lively than the two decentralized policies, we have included it in the comparison. 

An idea to use RCBFs to robustify the CCS policy against the missing information 
led to the development of the ``Predictor-Corrector for Collision Avoidance" (PCCA) method (see \cite{arXiv}):

\noindent
{\bf PCCA QP} (for agent $i$): Find control actions $u_{ij}, j=1,\ldots, N_a$ 
such that
\beq \barr{l}{\ds \min_{u_{i1}, \ldots, u_{iN_a}} \left(  \|u_{ii} -u_{0i}\|^2 +\sum_{j=1, j\not = i}^{N_a} \|u_{ij}\|^2 \right)\  {\rm subject \  to}  \\*[2mm]
 \ds a_{ij} + b_{ij}(u_{ii}- u_{ij}- \hat w_{ij})\ge 0  \ \forall  j =1,\ldots, N_a , j\not = i \\*[2mm]
 \ds a_{jk} + b_{jk}(u_{ij}+ \hat w_{ij} - u_{ik} - \hat w_{ik})\ge 0  \ \ \forall  j,k =1,\ldots, N_a \\*[2mm]
 \ds \hspace*{2.0in} j\not = k \ {\rm and} \ j,k \not = i}\label{rQPp} \eeq
and implement its own: $u_i = u_{ii}^*$.

This setup resembles the CCS (\ref{rQPccs}) with all the agent-to-agent constraints accounted for, but with no $\rho_i$
multiplier applied as in CCS. Instead, the (fictitious) disturbance terms $\hat w_{ij}$  have been added to the $u_{ij}$ ($i \not = j$). They represent the
uncertainty of agent $i$'s computation of agent's $j$ acceleration. One could put an upper limit on this uncertainty and
proceed with RCBF using the worst case disturbance. Instead, the PCCA uses the estimated disturbance as a difference
between the control action for agent $j$ ($u^*_{ij} $) computed by the host (agent 
$i$) with the action agent $j$ actually implemented ($u_j$):
\beq \hat w_{ij} = u_j - u^*_{ij} \label {hatwj} \eeq 
Because $u_{ij}^*$ requires knowing $\hat w_j$ and vice versa, an algebraic feedback loop is created.
To break this algebraic loop, one could use either the value from the previous sample (the controller solving
the QP could only be implemented in discrete time) or a low pass filter.

The paper \cite{arXiv} considered the case of two agents and, with discrete single sample delay, proved that the error in
enforcing the constraint is of the order of the sample time $\Delta T$ -- the smaller the sampling time, the smaller the error. 
Moreover, \cite{arXiv} showed that, even if one agent is not cooperating, 
the other agent takes over full responsibility for collision avoidance also producing possible constraint violation of the order of the sample 
time. In contrast to CCS, we could not find a multi-agent case 
when the real constraints would not be satisfied (within $\Delta T$ accuracy) but the general proof is not available. We also note that PCCA assumes 
information (measurement) of other agents' acceleration that can generally be obtained by lead filtering velocity measurements/estimates or with the use of an estimator such as the one proposed in \cite{ansari}.

We now show that the Centralized, CCS, and PCCA QP's are feasible.
To the best of our knowledge, this is a new result, but specific to the distance-based barrier function. 
The problem is nontrivial because there is a scenario
where we have more active QP constraints than the linearly independent (row) vectors
multiplying control inputs. 
This situation also prevents the standard approach to establishing Lipschitz continuity of optimal programs from being used \cite{morris}
(note: we are not implying the controller is not Lipschitz continuous). 
\\
\begin{proposition} \label{prop_feasibility} The Centralized QP (\ref{rQPc}), CCS QP (\ref{rQPccs}), and PCCA QP (\ref{rQPp}) are always feasible in the admissible set 
${\mathcal C}^* = \{ x \in {\R}^n: h_{ij}(x) \ge 0, h_{ij}(x) \ge -\frac{1}{\lambda_1}\dot h_{ij}(x), i,j \in\{1, \ldots, N_a\}, i\not = j \}$ and
the solution in each case is unique. 
\end{proposition}
\vspace{3mm}
\noindent
{\bf Proof}: \ 
Consider the Centralized policy constraint $F_{ij} := a_{ij} + b_{ij}(u_{i} - u_j)$. From the definition of $a_{ij}$, $b_{ij}$ we have
\[ F_{ij} \ge 2\|v_{ij}\|^2 + 2\xi_{ij}^T(u_i-u_j) + 2\lambda_1\xi_{ij}^T v_{ij} \]
where the last term is obtained by using
$h_{ij}\ge \frac{-1}{\lambda_1 }\dot h_{ij}$ (from the definition of ${\mathcal C}^*$) and
$l_1 - l_0/\lambda_1 = \lambda_1$.
From here, we construct a feasible $u$ by selecting one that satisfies 
\beq u_i - u_j = -\lambda_1 v_{ij}  \label{u12_ind} \eeq
which results in $F_{ij} \ge 2\|v_{ij}\|^2 \ge 0$. 

We proceed by using mathematical induction. 
For the first two agents, we pick any $u_1$ and $u_2$ that satisfy
(\ref{u12_ind}) 
where, in this case, $i = 1$ and $j=2$.
For example, we could select $u_1 = 0, u_2 = \lambda_1 v_{12}$.
Proceeding with the induction argument, assume that for the first $l-1$ agents we have selected 
control inputs $u_1, \ldots, u_{l-1}$ such that the condition (\ref{u12_ind}) 
 holds for all $i,j = 1,\ldots, l-1, \ i\not = j$. Adding the $l$-th agent
 we first consider $F_{1l} \ge 2\|v_{1l}\|^2 + 2\xi_{1l}^T(u_1-u_l +\lambda_1 v_{1l})  $.
Selecting $u_l = u_1 + \lambda_1 v_{1l}$ makes $F_{1l}\ge 0$ and we need to show that 
 all the other constraints are satisfied. Because $v_{il} = v_{ij} + v_{jl}$, $v_{ij} = -v_{ji}$, and 
 (\ref{u12_ind}) holds for $i,j = 1,\ldots, l-1, \ i\not = j$ by the induction assumption, 
 for all $i = 2, \ldots, l-1$ we have 
 \[ \barr{l}{ \lambda_1 v_{il} + u_i - u_l = \lambda_1 (v_{i1} + v_{1l}) + u_i - u_l \\*[2mm]
  \hspace*{6mm} = -u_i+u_1 + \lambda_1 v_{1l} + u_i - u_l = 0 }\]
Thus, for all $i = 1, \ldots, l-1$, $F_{il} \ge 2\|v_{il}\|^2 \ge 0$ and the induction argument completes the feasibility part for the 
centralized QP.  Because the optimal program has (strictly convex) quadratic cost and linear (convex) constraints, there is a unique solution. 
 
 Feasibility of CCS follows because, by changing the variables, the constraint set takes the same form as that of the centralized controller
 with only the cost function being different. The same applies for PCCA.
\hspace*{\fill} $\bigtriangledown$ \\
 
\begin{remark} 
From the proof of Proposition \ref{prop_feasibility} it is clear that we have one extra degree of freedom assuring feasibility 
even if one agent, say agent 1, is non-responsive, but with its acceleration known to the entity computing 
the QP. Second,  if we introduce a fictitious, stationary  agent 1, 
each agent braking proportional to its velocity
$u_i = -\lambda_1 v_i$ becomes a feasible action.\footnote{We note that the fixed (maximal) deceleration policy does not work for the barrier functions and constraints considered in this paper.
For example, with $\| \xi_{ij}\| ^2= r ^2/(1-\varepsilon)$ (with $\varepsilon < 1$), $v_i$ and $v_j$ collinear in the same direction, and $v_{ij} = 
-\lambda_1 (1-r^2/\|\xi_{ij}\|^2) \xi_{ij}$, $F_{ij}$ constraint would be violated if both agents decelerate with the same 
constant deceleration. }
 Note that  the proportional braking policy is a sub-optimal option
(it leaves $F_{ij} \ge 2\|v_{ij}\|^2$) proving feasibility, rather than an external action to be applied when the respective QP is not feasible.
The above consideration also shows that the deceleration for each agent need not be larger than $ \lambda_1 v_i$,
Thus, the QP problem remains feasible  even if agent deceleration is limited provided their speed is appropriately limited too. 
\end{remark}

\section{5-Agent Simulation Results}\label{sec:sims}
\noindent
We now compare the CBF collision-avoidance algorithms reviewed above by Monte-Carlo simulation of five agents maneuvering in an enclosed area. The agents are modeled as disks of radius $r_0 = 2$ with the center motion given by a double integrator in two dimensions as in \eqref{agent_i}. A static outer circle of radius $R_0 = 11$ acts as an additional (soft) barrier constraint to enclose the space containing all five agents. The controller sample time is chosen to be $\Delta T = $50 ms, and the baseline controller $u_{0i}$ for each agent is computed by LQR with $Q=0.2I_4$ and $R=I_2$. For computation of the QP constraints \eqref{h_constr1}, we choose $l_0 = 6$ and $l_1 = 5$ to satisfy $l_1^2 \ge 4l_0$ (i.e. negative real eigenvalues). 
All the algorithms use this same set of parameters and CCS is implemented with $\rho_i = 2$. 

\begin{figure}[t!]\vspace{-.0in}
    \centering
\includegraphics[scale=0.78]{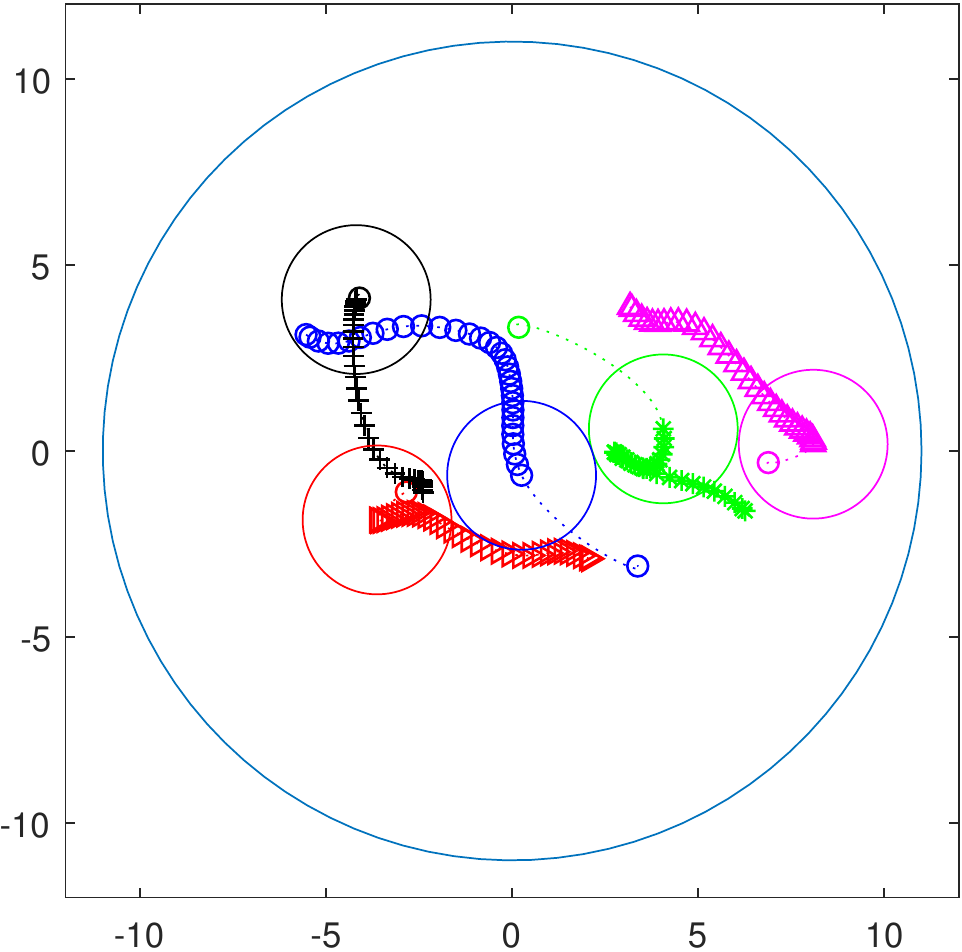}
    \vspace{-0.1in}
    \caption{5-agent simulation time snapshot depicting past and future paths. Each agent has a randomly defined non-conflicting beginning/end location.}
    \label{fig:5agent}
\end{figure}

For each simulation trial, each agent is assigned random beginning and end locations somewhere within the outer static circle area. These initial/final locations are assessed for any agent-to-agent overlap as well as overlap with the outer static circle. If a physical overlap is indicated, new beginning and/or end locations are assigned until 100 feasible initial and final positions are ensured. Each algorithm then ran from the same 100 feasible initial positions to the corresponding final positions. Figure \ref{fig:5agent} shows a time snapshot of one of these rans -- the agents, their beginning and end locations, as well as their past and future paths are all displayed.

To assess the algorithms, a set of metrics was devised to compare liveness, collisions, and feasibility. Liveness is a measure of convergence time; we assess how long it takes for all the agents to reach within 0.1 units from their destinations with the velocity magnitude less than 0.1 units/sec. Each simulation was run for 100 seconds and assessed for convergence. It was found that all non-convergent runs at 100 seconds had gridlocked and were not expected to converge.
We did not use the deconfliction algorithms for gridlocks because they need a preferred passing direction to be agreed up front \cite{wang} or determined on line, which
assumed agent-to-agent communication in \cite{celi}.

The results shown in Table \ref{tab:compare} depict the aggregated results from 100 Monte-Carlo simulations for the Centralized \eqref{rQPc}, DF \eqref{rQPd}, DR \eqref{rQPr}, CCS \eqref{rQPccs}, and PCCA \eqref{rQPp} policies without any additional radius margin added. PCCA was implemented with either a sample delay ($\Delta T = 50$ms) or a low-pass filter with a time constant of 0.2 sec to break the algebraic loop between (\ref{rQPp}) and \eqref{hatwj}. Table \ref{tab:compare} shows the minimum convergence time to be similar for each algorithm,
%
%
\begin{table}[b!]\vspace{-0.2in}
	\caption{Metrics for CBF-based collision-avoidance algorithms from 100 Monte-Carlo simulation runs with no radius margin}	
	\begin{tabular}{|l|ccc|c|c|c|}
	\cline{1-7}
	                                & \multicolumn{3}{c|}{Converge Time (sec)}             &                & \#     & \#         \\
	                Method                & min                   & max                   & mean & $h_{\rm{min}}$ & gridlock & infeasible \\ \hline
	\multicolumn{1}{|l|}{Centralized}      & \multicolumn{1}{c|}{7.45} & \multicolumn{1}{c|}{22.15} & 12.98    & -0.002    & 0         & 0          \\ \hline
	\multicolumn{1}{|l|}{DF}           & \multicolumn{1}{c|}{7.55} & \multicolumn{1}{c|}{67.20} & 17.44    & -2.84      & 3         & 27        \\ \hline
	\multicolumn{1}{|l|}{DR}           & \multicolumn{1}{c|}{7.55} & \multicolumn{1}{c|}{84.75} & 17.26    & -1.53      & 4         & 32        \\ \hline
	\multicolumn{1}{|l|}{CCS$_{2}$}         & \multicolumn{1}{c|}{7.60} & \multicolumn{1}{c|}{31.25} & 14.63    & -1.35      & 4         & 0          \\ \hline
	\multicolumn{1}{|l|}{PCCA}         & \multicolumn{1}{c|}{7.35} & \multicolumn{1}{c|}{23.75} & 12.76    & -0.015      & 0         & 0          \\ \hline
	\multicolumn{1}{|l|}{PCCA$_{0.2}$} & \multicolumn{1}{c|}{7.35} & \multicolumn{1}{c|}{21.65} & 12.68    & -0.067    & 0         & 0          \\ \hline
	\end{tabular}
	\vspace{-0.1in}
	\label{tab:compare}
\end{table}
%
\begin{figure}[t!]\vspace{-.0in}
    \centering
\includegraphics[scale=0.66]{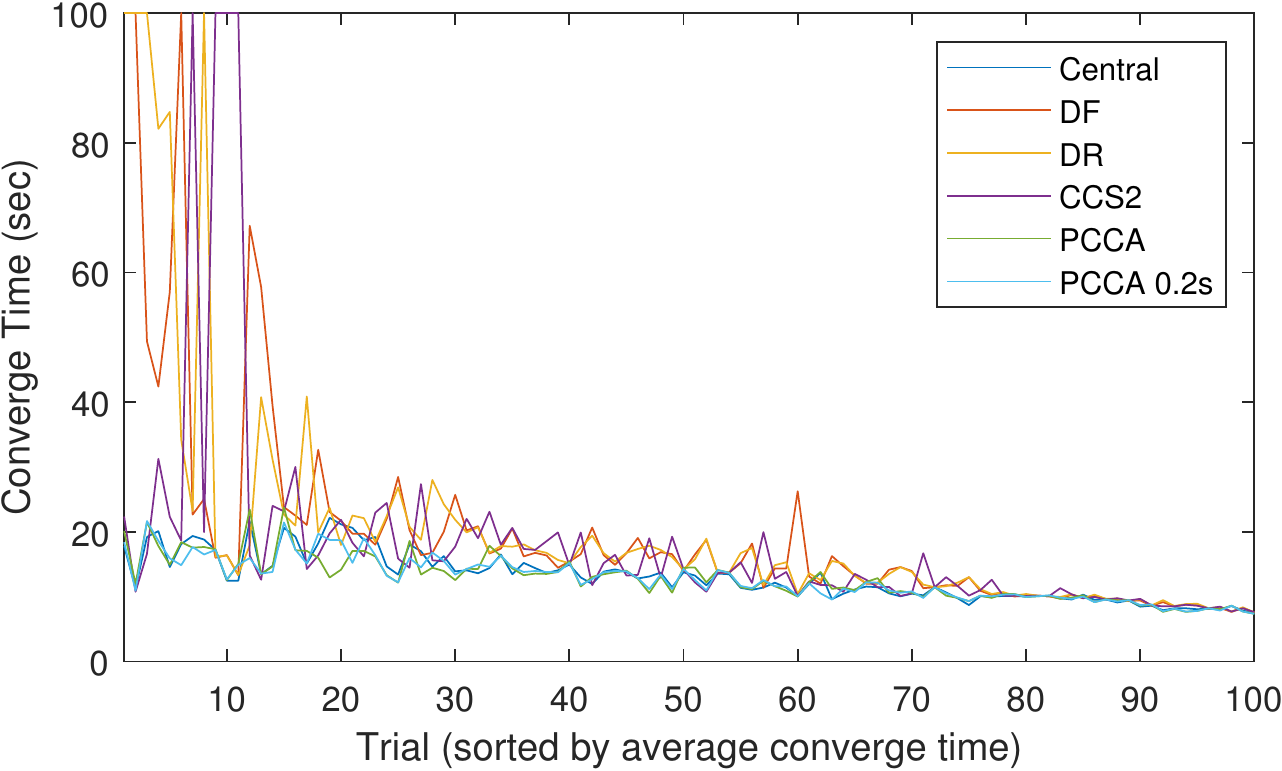}
    \vspace{-0.1in}
    \caption{Convergence time for 100 Monte Carlo simulation runs, sorted by average from max to min excluding gridlocked trials.}
    \label{fig:time}
\end{figure}
%
%
\begin{figure}[t!]
    \centering
\includegraphics[scale=0.66]{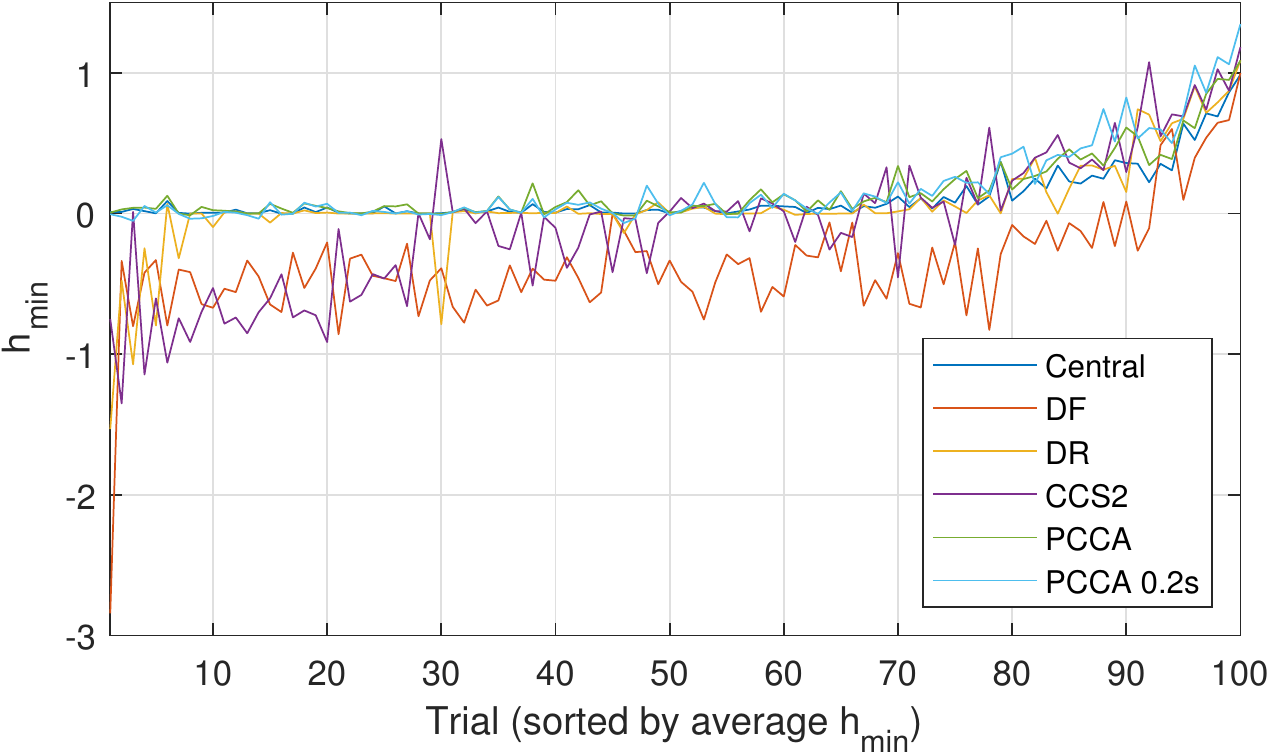}
    \vspace{-0.1in}
    \caption{Minimum barrier distance for 100 5-agent Monte Carlo simulations, sorted by average from min to max.}
    \vspace{-0.1in}
    \label{fig:hmin}
\end{figure}
%
while the maximum is quite varied. Additionally, the min, max, and mean values do not include the gridlocked simulation results for DF, DR, and CCS. Both the DF and DR algorithms exhibit less liveness and generally take longer for all agents to converge as previously reported in the literature (e.g. \cite{wang}). We see this better in Figure \ref{fig:time}, where the simulations trials are sorted by the average convergence time for each trial over all algorithms from maximum to minimum. Generally, the convergence times and liveness of the Centralized and PCCA algorithms are similar, while DF, DR, and CCS exhibit longer times to converge, if at all.

It has been established above that the centralized, CCS, and PCCA controllers are always feasible and the simulations confirmed this. 
However, nearly a third of the DF and DR simulations exhibited infeasible QPs at some point. In this case, the QP solver \cite{odys} was configured to return the "least infeasible" solution before implementing the control. Except for slacked constraints on the outer static circle, the algorithms were implemented in pure form without alternative actions to handle infeasibility.

For collision avoidance in the multi-agent case, the centralized controller exhibits the best results, but also requires explicit communication. While the barrier is shown to be violated (i.e. $h_{min} = -0.002$, $h_{min}$ is the minimum agent-to-agent barrier value over all the agents and all 100 trials), this is due to the selection of sampling time. When the sampling time was reduced, this barrier violation disappeared as expected. Both the PCCA controller with unit delay and the PCCA controller with low-pass filter perform almost as well as the centralized controller.  A pictorial comparison is shown in Figure \ref{fig:hmin} that displays the minimum agent-to-agent barrier value sorted by average  of $h_{min}$ over all the algorithms. DF and CCS generally have larger violations than the other methods while DR has only a few visible violations. Both the Centralized and PCCA controllers exhibit minimal barrier violation and are almost indistinguishable in the plot.

\begin{figure}[htbp!]\vspace{-.1in}
\hspace{-0.15in}\includegraphics[scale=0.64]{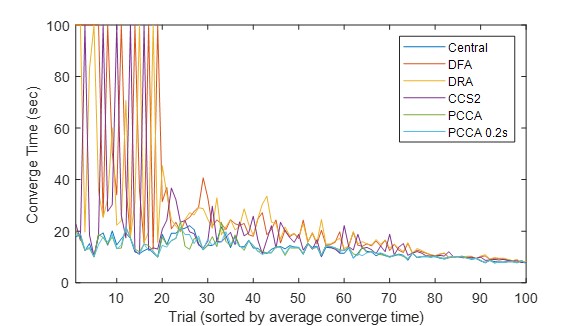}
    \vspace{-0.13in}
    \caption{Convergence time for 100 Monte Carlo simulation runs with added margin, sorted by average from max to min excluding gridlocked trials.}
    \label{fig:time2}
     \vspace{-0.25in}
\end{figure}

\begin{figure}[t!]\vspace{-.1in}
\hspace{-0.18in}\includegraphics[scale=0.64]{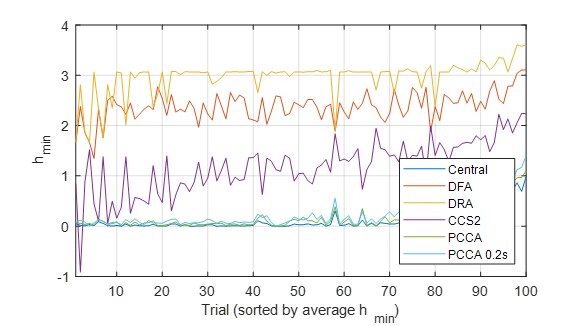}
    \vspace{-0.14in}
    \caption{Minimum barrier distance for 100 5-agent Monte Carlo simulations with added margin, sorted by average from min to max.}
    \vspace{-0.1in}
    \label{fig:hmin2}
\end{figure}

We now rerun the Monte-Carlo simulation trials using the worst-case agent-to-agent barrier violations $h_{min}$ recorded in Table \ref{tab:compare} to add a radius margin ($r > 2 r_0$) for each algorithm's computation of the barrier constraint $h$. The results, shown in Figures \ref{fig:time2} and \ref{fig:hmin2} and tabulated in Table \ref{tab:compare_marg}, depict the barrier violations with the agents' actual size $r_0$; all methods but one CCS trial effectively avoid collision with radius margin added. The CCS trial with a collision is due to an interaction between agent-to-agent hard constraints and agent-to-static-outer-circle soft constraints. Note also that the radius margins added for DF and DR grow the agent sizes enough to induce additional gridlocks and infeasibility. One could iterate on the barrier margin required for both DF and DR to achieve $h_{min}$ closer to zero, similar to the other methods.

%
%
%
%

%
\begin{table}[h!]\vspace{-0.01in}
	\caption{Metrics for CBF-based collision-avoidance algorithms from 100 Monte-Carlo runs after adding worst-case agent-to-agent violation to radius margin.}	
	\begin{tabular}{|l|ccc|c|c|c|}
	\cline{1-7}
	                                & \multicolumn{3}{c|}{Converge Time (sec)}             &                & \#     & \#         \\
	                 Method   & min                   & max                   & mean & $h_{\rm{min}}$ & gridlock & infeasible \\ \hline
	\multicolumn{1}{|l|}{Central}      & \multicolumn{1}{c|}{7.45} & \multicolumn{1}{c|}{22.15} & 12.98    & 0.000    & 0         & 0          \\ \hline
	\multicolumn{1}{|l|}{DF}           & \multicolumn{1}{c|}{7.90} & \multicolumn{1}{c|}{50.60} & 18.03    & 1.34      & 11         & 37        \\ \hline
	\multicolumn{1}{|l|}{DR}           & \multicolumn{1}{c|}{7.95} & \multicolumn{1}{c|}{82.95} & 19.91    & 1.67      & 5         & 38        \\ \hline
	\multicolumn{1}{|l|}{CCS$_{2}$}         & \multicolumn{1}{c|}{7.80} & \multicolumn{1}{c|}{36.75} & 15.37    & -0.91      & 5         & 0          \\ \hline
	\multicolumn{1}{|l|}{PCCA}         & \multicolumn{1}{c|}{7.35} & \multicolumn{1}{c|}{23.75} & 12.77    & -0.002      & 0         & 0          \\ \hline
	\multicolumn{1}{|l|}{PCCA$_{0.2}$} & \multicolumn{1}{c|}{7.35} & \multicolumn{1}{c|}{21.80} & 12.70    & 0.001    & 0         & 0          \\ \hline
	\end{tabular}
	\vspace{-0.1in}
	\label{tab:compare_marg}
\end{table}
%

While CCS did deliver infeasibility-free runs, as expected, the collisions and the number of gridlocks was similar to the pure decentralized policies DF and DR. Another surprising finding was the very close, practically indistinguishable results for the Central and the PCCA policies despite the fact that the former runs with full information, while the latter does not know where the other agents are heading, i.e. does not know their $u_{0j}$. Next, we will explain these results on a related problem simplified enough to make it analytically tractable.

\section{Equilibrium instability and liveness} \label{sec:instability}
\noindent
In this section we  consider a simpler
problem with only two agents, but to make the problem non-trivial for our objective, we give them only one degree of freedom -- they can only 
accelerate or decelerate. The problem is equivalent to robots or vehicles moving through narrow  straight corridors where turning is 
not a useful option. Essentially, the agents will have to decide (negotiate) which one goes first through the intersection. 
It  turns out that the policies that exhibited gridlocks (DF, DR, CCS) create a set of stable equilibrium 
(gridlock) points, while for those that did not (Centralized, PCCA) there is a single unstable equilibrium point in the physical space.
In other words, the liveness in these multi-agent systems comes from instability. 

\begin{figure}[htbp!]\vspace{-.05in}
    \centering
\includegraphics[scale=0.66]{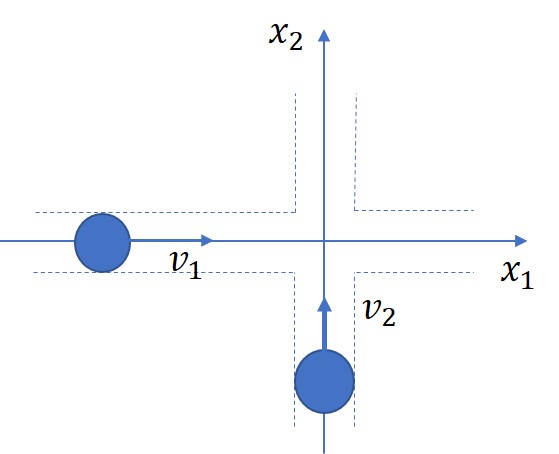}
    \vspace{-0.1in}
    \caption{The configuration for the two agents inside their corridors -- the corridor lines just serve for illustration, the agents cannot turn.}
    \vspace{-0.1in}
    \label{fig:two-agents}
\end{figure}

The problem considered here is depicted in Figure \ref{fig:two-agents}. Each agent can only move longitudinally,  accelerate or decelerate. 
The control action is the velocity, i.e. the model of each agent is the single integrator. For agent $i$, $i = 1,2$, the model is
\beq \dot x_i = v_i \label{agent_i2} \eeq
Here, we have used $x_1$ and $x_2$  (instead of $y_2$) to denote the distance of the two agents from the origin -- the intersection point. The goal of each agent is to move at their desired velocity $v_{0i}$, assumed constant, while avoiding collision with the other agent. We assume that the initial displacements 
from the origin are negative while the desired velocities are positive as depicted in Figure \ref{fig:two-agents}. For collision avoidance, we use 
the same distance-based control barrier function 
\beq h(x) = x_1^2 +x_2^2 - r^2 = x^Tx - r^2  \label{h} \eeq
with $r\ge 2r_0$, $x = [x_1,x_2]^{\rm T}$, and the admissible set ${\mathcal C} = \{x\in {\R^2}: h(x) \ge 0\}$. The barrier constraint is therefore given by 
\beq   2x_1 v_1 + 2x_2 v_2 + \lambda h(x_1, x_2)  \ge 0 \label{h_constr} \eeq
where $\lambda> 0$ is a design parameter -- the barrier bandwidth. 
Because $x_1$ and $x_2$ cannot be simultaneously 0 in the admissible set ${\mathcal C}$, if both control actions $v_1$ and $v_2$ are available for collision
avoidance, the problem is feasible and $h$ is a CBF. 
   			
\subsection{DF and DR equilibria and their stability}
\noindent
For the DR policy, each agent assumes half of the responsibility for satisfying the barrier constraint while having only its own 
control at its disposal. For agent $1$, the QP is
\beq \barr{l} {\min_{v_1} \|v_1 - v_{01}\|^2  \ \ {\rm such \ that} \\*[2mm]
          \frac{\lambda}{2} h + 2x_1 v_1 \ge 0 } \label{dr_a1} \eeq
Note that the only difference between the DR and DF policies would amount to removing $\frac{1}{2}$, or equivalently selecting
$\lambda$ larger by a factor of 2. Hence, all the conclusions here apply to DF as well. The solution to (\ref{dr_a1}) for agent 1 is given by
\beq v_1^* = \left\{ \barr{ccl}{ v_{01} & if & \frac{\lambda}{2} h + 2x_1 v_{01} \ge 0 \\*[2mm]
- \frac{\lambda}{4x_1} h  & if & \frac{\lambda}{2} h  + 2x_1 v_{01}  < 0 } \right . 
\label{v_star_df} \eeq
The case for agent $2$ is symmetric: the subscript $``1" \rightarrow ``2"$.  

\begin{figure}[htbp!]\vspace{-.05in}
    \centering
\includegraphics[scale=0.75]{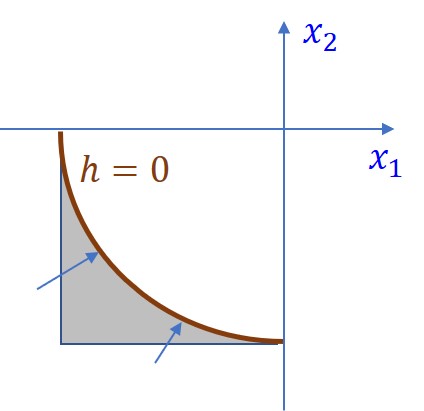}
    \vspace{-0.1in}
    \caption{The equilibrium set for the Decentralized Follower policy.}
    \vspace{-0.05in}
    \label{fig:dr_equil}
\end{figure}

Now, we consider the critical case, in terms of gridlocks, where the barrier constraints for both agents are active:
\beq \barr{l}{ \dot x_1 = - \frac{\lambda}{4x_1} h \\*[2mm]
  \dot x_2 = - \frac{\lambda}{4x_2} h     }  \label{dr_dynamic} \eeq
 First, because $x_i < 0$,  both agents' constraints will activate while $h(x) >0$. After
  they activate, $\dot h = -\frac{\lambda}{2} h$, which results in $h(x(t)) > 0 \ \forall t$ and guarantees collision avoidance.
  Second, the only way for the two agents satisfying (\ref{dr_dynamic}) to move backward is if $h(x) < 0$, but this has just been ruled out. 
Therefore, once they both get closer to the origin than
  $r$ (the shaded region in Figure \ref{fig:dr_equil}) they are stuck because neither moves backward and
  one of them has to be farther away than $r$ for the other to pass. Thus, any trajectory that 
  hits the shaded ``triangle" will result in a gridlock. 

  More formally, we look at the equilibrium structure of (\ref{dr_dynamic})
  in the quarter-plane $\{ x_1 < 0, x_2 < 0\}$ where $h(x) = 0$. Thus, the equilibrium set 
  is an arc or radius $r$ as shown in Figure \ref{fig:dr_equil}. Moreover, because
  $\dot h = - \frac{\lambda}{2} h$, the arc itself is attractive from the points outside of it. Linearizing around an equilibrium point $x_e \in\{x\in {\R^2}:
  h(x) = 0,x_1 < 0, x_2 < 0\}$,
  we obtain the linear system  $\dot \zeta = A_{dr} \zeta$ in the $\zeta = x - x_e$ coordinates, where 
  \beq A_{dr}  = \mx{cc}{ - \frac{\lambda}{2} & - \frac{\lambda x_{2e}}{2x_{1e}} \\*[2mm] 
  - \frac{\lambda x_{1e} }{2 x_{2e} } & - \frac{\lambda}{2} } \label{A_dr} \eeq
 The eigenvalues of the matrix $A_{dr}$ are $0$ and $- {\lambda}$. This does not necessarily prove that the equilibrium $x_e$
 is stable because the system (\ref{dr_dynamic}) is nonlinear. Instead, we consider $g(x) =x_1^2 - x_2^2$ which satisfies $\dot g =0$.
 Hence, the function $g(x)$ is constant along the trajectories of (\ref{dr_dynamic}) and intersects 
 the arc $h(x)=0$ if $|g(x(t_c))| < r^2$ in the third quadrant. Here,
 $t_c$ is the first time instant when both agents have their constraints activated. 
 This shows that the points on the equilibrium arc are indeed stable and that a small 
 perturbation or noise would not result in them clearing the intersection.

\subsection{Centralized controller equilibrium analysis} 
\begin{figure}[htbp!]\vspace{-.05in}
    \centering
\includegraphics[scale=0.65]{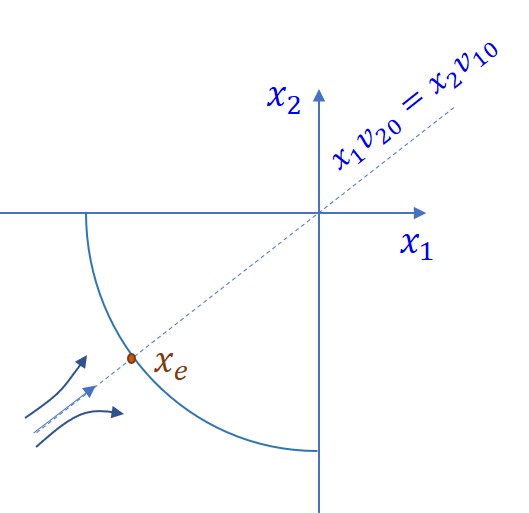}
    \vspace{-0.1in}
    \caption{The equilibrium set for the Centralized policy.}
    \vspace{-0.05in}
    \label{fig:central}
\end{figure}

\noindent
With the Centralized controller, the agents have full information about each others intentions and solve the same
QP:
\beq \barr{l} {\min_{v_1, v_2} \|v_1 - v_{01}\|^2  + \|v_2 - v_{02}\|^2 \ \ {\rm such\ that}   \\*[2mm]
          \lambda h + 2x_1 v_1 + 2x_2 v_2 \ge 0 } \label{central} \eeq
 This QP is executed by both agents and they would activate the constraint at the same time, namely when 
 $ \lambda h + 2x_1 v_{01} + 2x_2 v_{02} $ becomes negative. Before this point, each agent uses 
 its desired constant velocity $v_{0i}$ and they move in the straight line in the $(x_1, x_2)$ space. After this point, the 
 coupled dynamics becomes
 \beq \barr{l}{ \ds \dot x_1 = - \frac{\lambda x_1}{2\|x\|^2} h  + \frac{x_2^2}{\|x\|^2} v_{01} -\frac{x_1 x_2}{\|x\|^2} v_{02}  \\*[3mm]
  \ds \dot x_2 = - \frac{\lambda x_2}{2\|x\|^2} h  - \frac{x_1 x_2}{\|x\|^2} v_{01}   + \frac{x_1^2}{\|x\|^2} v_{02}    }  \label{central_dynamic} \eeq
By adding and subtracting the right hand sides of (\ref{central_dynamic}) and setting them to 0 we obtain the equilibrium set, defined by the intersection
of the arc $h(x) = 0$ and the line $x_2 v_{01 }- x_1 v_{02} = 0$ as shown in Figure \ref{fig:central} -- that is, the equilibrium is a single point.

Linearizing around the equilibrium, we obtain the dynamics $\dot \zeta = A_c \zeta$, with the state $\zeta = x-x_e$ as above, and the matrix $A_c$
given by
  \beq A_{c}  = \mx{cc}{\ds  - \frac{\lambda x_{1e}^2 }{\|x_e\|^2}  - \frac{x_{2e} v_{02} }{\|x_e\|^2} & \ds - \frac{\lambda x_{1e} x_{2e} }{\|x_e\|^2} + 
   \frac{x_{2e} v_{01} }{\|x_e\|^2}\\*[2mm] 
 \ds  - \frac{\lambda x_{1e} x_{2e} }{\|x_e\|^2} + \frac{x_{1e} v_{02} }{\|x_e\|^2} & \ds  - \frac{\lambda x_{1e}^2 }{\|x_e\|^2} 
   - \frac{x_{1e} v_{01} }{\|x_e\|^2}  }\label{A_c} \eeq
The eigenvalues of the matrix $A_c$ are 
\[ {\rm eig(}A_c) = \left \{ -\lambda, \frac{\sqrt{v_{01}^2 + v_{02}^2}}{r} \right \} \]
where we have used the solution for the equilibrium: $x_{ie} = -\frac{v_{0i} r}{\sqrt{v_{01}^2 + v_{02}^2}}, \ i = 1,2$. 
Note that the second eigenvalue is positive and the equilibrium point $x_e$ is unstable, explaining 
the lack of gridlocks for the Central policy. Only the trajectories that start exactly on the stable manifold -- a set (in this case a line) of measure 0 -- would end up at the equilibrium (that is, gridlocked), while all other trajectories move exponentially fast away allowing agents to clear the intersection.

\subsection{CCS equilibrium analysis}
As before, we give CCS the control over both agents while realizing that only the host action computed by the QP 
is the actual one (for agent 1, $v_1 = v_{11}^*$) while the other one ($v_{12}^*$)  is virtual. We assume no information about the other agent's intention, 
so the agent 1 QP uses $v_{02} = 0$, and vice-versa. Upon changing variables, $\bar v_{11} = v_{11} - v_{01}$, the QP problem for the agent 1 is
\beq \barr{l} {\min_{\bar v_{11}, v_{12}} \|\bar v_{11} \|^2  + \|v_{12} \|^2 \ \ {\rm such\ that}   \\*[2mm]
          \lambda h + \rho_1(x) 2x_1 v_{01} + 2x_1 \bar v_{11} + 2x_2 v_{12} \ge 0 } \label{CCS} \eeq
In this case, the factors $\rho_i$'s are used to assure that, when the constraint is active for both agents, the actual 
barrier constraint (\ref{h_constr1}), the one that guarantees collision avoidance, is satisfied. The control computed by agent 1 for itself (the other one it computes gets discarded) is given by 
\[ v_{1} = v_{01} - \frac{\lambda h x_1 + \rho_1(x) 2 x_1^2 v_{01}}{2\|x\|^2} \]
To decide what $\rho_i$ to use, we look at the CBF constraint $\dot h + \lambda h\ge 0$ when each agent implements its own control:
 $\lambda h +2x_1 v_{1} + 2 x_2 v_{2} = \sum_{i=1}^2 2 x_i v_{0i}(\|x\|^2 - \rho_i(x) x_i^2) \ge 0$. 
Because the two terms in the sum are independently computed, each has to be non-negative. A good choice then is
$\rho_i = 1 + \frac{x_j^2}{x_i^2} $, $i$ being the host, $j$ the other agent. 
This choice of $\rho_i$'s produces the joint dynamics under active constraints as follows:
\beq \barr{l}{ \dot x_1 = - \frac{\lambda h }{2\|x\|^2} x_1 \\*[2mm]
  \dot x_2 = - \frac{\lambda h }{2 \|x\|^2} x_2  }  \label{ccs_dynamic} \eeq
Thus, a part of the equilibrium analysis is the same as in the DR/DF case: 
\begin{itemize}
\item The equilibrium set, given by $h(x) = 0$, is attractive, i.e. $\dot h = -\lambda h$, guaranteeing there are no collisions. 
\item  The linearization shows the eigenvalues at any equilibrium point $x_e$ to be ${\rm eig} (A_{ccs}) = \{0, -\lambda\}$. 
\item The function invariant along the trajectories of
(\ref{ccs_dynamic}) is $g_{ccs} = \ln(-x_1) - \ln(-x_2) + c$. 
\end{itemize}
Thus, the arc $h(x) =0$ represent a set of stable (gridlock) equilibrium points. 
However, the presence the $x_i^2$ in the denominator in $\rho_i$ creates an additional (spurious) 
equilibrium of (\ref{ccs_dynamic}) at $x_i = 0$ not related to the position of the other agent. 
This singularity in $\rho_i$ can be removed by multiplying both side of (\ref{CCS}) by $x_i^2$, but the equilibrium remains. 
\begin{itemize}
\item Selecting $\rho_i = 1 + \frac{x_j^2}{x_i^2} $ makes the constraint in (\ref{CCS}) active when $x_i$ is negative and sufficiently close to 0. 
\item Making $\rho_i$ smaller than $1 + \frac{x_j^2}{x_i^2} $ leads to  a violation of the actual constraint, while making it larger makes the agents stop before they arrive at  the original equilibria $h = 0$ or $x_i = 0$.
\end{itemize}
Hence, the  CCS policy always produces a gridlock either at $h = 0$ or $x_i = 0$. The first one will not be impacted by a small amount of noise. A small amount of noise would eventually free the agents from the second one -- once $x_i > 0$ due to noise, the constraint becomes inactive and $\dot x_i = v_{i0}$.

\subsection{PCCA equilibrium analysis} 
\noindent
As was the case with the CCS policy, the PCCA policy performs co-optimization and uses $0$ instead of the unknown desired velocity for
the other agent. The ``error," i.e. the difference between the observed and host computed velocity for the other agent, 
is filtered and fed back as the 
disturbance term into the constraint:
\beq \barr{c} {\! \! [v_{11}^*, v_{12}^*] \! = \! \arg \min_{v_{11}, v_{12}} \|v_{11} - v_{01}\|^2  + \|v_{12} \|^2 \ {\rm such\ that}   \\*[2mm]
          \lambda h + 2x_1 v_{11} + 2x_2 (v_{12} + w_2)\ge 0  \\*[2mm]
       {\rm where} \ \    \dot w_2 = \frac{1}{\tau} (-w_2 + v_2 -v_{12}^*) } \label{PCCA} \eeq
The closed form solution to the PCCA QP is given by
 \beq [v_{11}^*, v_{12}^*]= \left\{ \barr{ccl}{ [v_{01}, 0] & if & \mu_1 \ge 0 \\*[2mm]
[v_{01}, 0] -\frac{\mu_1}{2\|x\|^2} x^T & if & \mu_1 < 0 } \right . 
 \label{v_star_pcca} \eeq
 with $\mu_1 = \lambda h + 2x_1 v_{01}+2x_2 w_2$.
The control $v_1 = v_{11}^*$ is then implemented as the velocity for agent 1.   The equations for agent 2 are symmetric.
  
 If both agents have active constraints, which is the only way to have a gridlock because inactive constraints lead to
 $ v_i = v_{0i} > 0$, the dynamics for the combined system is given by 
 \beq \barr{l}{ \dot x_1 = -\frac{\lambda h}{2\|x\|^2} x_1 + \frac{x_2 v_{10} - x_1 w_2}{\|x\|^2} x_2 \\*[2mm]
 \dot x_2 = -\frac{\lambda h}{2\|x\|^2} x_2 - \frac{x_2 w_1 - x_1 v_{20} }{\|x\|^2 }x_1 \\*[2mm]
\tau  \dot w_1 = \frac{x_2}{\|x\|^2}(x_2 v_{10}+x_1 v_{20} - x_2 w_1 - x_1 w_2) \\*[2mm]
\tau  \dot w_2 = \frac{x_1}{\|x\|^2}(x_2 v_{10}+x_1 v_{20} - x_2 w_1 - x_1 w_2) } \label{PCCA_dynamic} \eeq

Note that we should use a fast filter (i.e. the time constant $\tau$ small) because the constraint adherence is within
 ${\mathcal O}(\tau)$ margin of error. This could be established by using the singular perturbation argument (see Chapter 11 in \cite{khalil}).
 To briefly illustrate the mechanism behind this property, we note that, regardless of the combination of active constraints between the two agents, 
 the variable $z = x_2 w_1 + x_1 w_2$ always satisfies
 \beq \tau \dot z = -z + (x_1 v_{02} + x_2 v_{01}) + {\mathcal O}(\tau)\label{z_dyn} \eeq	
 As $\tau \rightarrow 0$, the solution of
 (\ref{z_dyn}) satisfies $z(t) = x_1(t) v_{02} + x_2(t)  v_{01} +  {\mathcal O}(\tau)$ for all $t \in[ t_0 + \delta, T]$, with $t_0$ the initial time, $\delta$ arbitrarily small, and $T$ arbitrarily large.
When both agents have active constraints, from (\ref{PCCA_dynamic}) we compute  
 \[ \dot h = -\lambda h + \frac{2x_1 x_2}{\|x\|^2} (x_2 v_{01} + x_1 v_{02} - z). \]
 In turn, 
 \beq h(t) = h(t_c) e^{-\lambda(t-t_c)} +  {\mathcal O}(\tau) \label{h_approx} \eeq
 proving that the barrier violation could be made arbitrarily small by selection of the time constant $\tau$.  
 The same equation (\ref{h_approx}) can be derived  for the other cases when either one or both agents have the constraint inactive by subtracting the inactive constraint(s) from the right hand side of $\dot h$ and using $z(t) = x_1(t) v_{02} + x_2(t)  v_{01} +  {\mathcal O}(\tau)$ (details omitted).  Hence, the barrier constraint is satisfied with ${\mathcal O}(\tau)$ error margin over any arbitrary long time interval $(t_0, T)$.

Returning to the equilibrium analysis of (\ref{PCCA_dynamic}), we obtain that the one-dimensional equilibrium set in $\R^4$ is defined by
\beq \barr{l}{ h(x_e) = x_{1e}^2 + x_{2e}^2 - r^2 =0  \\*[2mm]
     x_{1e} w_{2e} = x_{2e} v_{01} \\*[2mm]
     x_{2e} w_{1e} = x_{1e} v_{02} } \label{PCCA_equil} \eeq
 To analyze stability of an equilibrium point $(x_e, w_e)$ that belongs to the set defined by (\ref{PCCA_equil}), we 
 use the ratio $\mu_e = \frac{x_{1e}}{x_{2e}}$ and change the variables:
 \beq \barr{l}{ \eta_1 = x_1 - x_{1e} - \mu_e (x_2 - x_{2e}) \\*[2mm]
     \eta_2 = \mu_e(x_1 - x_{1e} ) + x_2 - x_{2e}   \\*[2mm]
     \eta_1 = \mu_e (w_1 - x_{1e}) - (w_2 - w_{2e}) \\*[2mm]
     \eta_4 = w_1 - w_{1e} + \mu_e (w_2 - w_{2e}) \\*[2mm]} \label{ch_of_var} \eeq
   The linearized system around $(x_e, w_e)$ in the $\eta$-coordinates is $\dot \eta = A_p \eta$ with $A_p$ given by
    \beq A_{p}  = \mx{cccc}{   \frac{v_{10} + \mu_e^3 v_ {20}}{r \mu_e \sqrt{1+\mu_e^2}}  & 0 & - \frac{\mu_e} {\sqrt{1+\mu_e^2} } & 0    \\*[2mm] 
 \frac{v_{10} - \mu_e v_ {20}}{r \mu_e \sqrt{1+\mu_e^2}}  & -\lambda & 0 & - \frac{\mu_e}{\sqrt{1+\mu_e^2} }   \\*[2mm] 
 0 & 0 & 0 & 0  \\*[2mm] 
\!   (\frac{v_{10}}{\mu_e}  - v_ {20}) \sqrt{1+\mu_e^2}  & 0 &  0  & - \frac{1} {\tau}    
     }\label{A_p} \eeq
where we have used $x_{1e} = -\mu_e r /\sqrt{1+\mu_e}$ and $x_{2e} = - r /\sqrt{1+\mu_e}$. The system $\dot \eta = A_p \eta$ is 
exponentially unstable because 
\[ {\rm eig}(A_p) = \left \{ \frac{v_{10} + \mu_e^3 v_ {20}}{r \mu_e \sqrt{1+\mu_e^2}}, -\lambda, 0, -\frac{1}{\tau}  \right \} \]
with the first eigenvalue being positive. Note that the 0 eigenvalue has moved from the physical states $x$ to the controller states $w$, while
the exponentially unstable state $\eta_1 = x_1 - x_{1e} - \mu_e (x_2 - x_{2e})$ belongs to the physical $(x_1, x_2)$ plane. The projection
of trajectories onto the physical space $(x_1, x_2)$ should resemble behavior 
shown in Figure \ref{fig:central} and helps explain the lack of gridlock for the 
PCCA policy in the simulations in Section \ref{sec:sims}. 
 
 \subsection{Summary of the equilibrium analysis and simulations}
 
 The analysis in this section correlates closely with the observed simulation results from Section \ref{sec:sims}. 
 While one distinction between decentralized, host-only-control policies 
and co-optimization policies is related to feasibility, it is not correlated to liveness and
 absence of gridlocks. The actual mechanism for gridlock avoidance is the presence of unstable equilibria in the joint agent space. 
 The policies that have unstable equilibria (Centralized and PCCA) produced no gridlocks in the 5-agent Monte Carlo simulations. 
 In contrast, the policies with stable equilibria had gridlocks observed. Existence of equilibria themselves cannot be avoided with the non-convex, compact obstacles as one could deduce from the results of \cite{braun}, Due to the particular
 structural difference of the 1-D problem, the CCS policy produces an extra equilibrium at the origin and, as a result, would always gridlock.
 The analysis of the equilibria structure and their stability is summarized in Table \ref{tab:equil}.\footnote{For CCS, the stable equilibrium 
may gridlock for some initial conditions. The equilibrium that always theoretically gridlocks is the spurious one at 0 that is not stable. A small amount of noise would eventually free the agent. }             
 
\begin{table}[h!]\vspace{-0.01in}
	\caption{Properties of the collision avoidance policies based on equilibrium analysis.}	
\hspace*{5mm}	\begin{tabular}{|l|c|c|c|c|c|c|}
	\cline{1-5}
	                                &  Target           &                        &  Equilibria             &      \\
	                 Method   &  Intent                  & Equilibria                 & Stability &            Gridlocks  \\ \hline
	\multicolumn{1}{|l|}{DR/DF}           & \multicolumn{1}{c|}{No} & \multicolumn{1}{c|}{Arc} & Stable    &  possible           \\ \hline
	\multicolumn{1}{|l|}{CCS}         & \multicolumn{1}{c|}{No} & \multicolumn{1}{c|}{Arc/axes} & Stable   & always
\\ \hline
	\multicolumn{1}{|l|}{Centralized}      & \multicolumn{1}{c|}{Yes} & \multicolumn{1}{c|}{Point} & Unstable   & measure 0         \\ \hline
	\multicolumn{1}{|l|}{PCCA$_{\tau}$} & \multicolumn{1}{c|}{No} & \multicolumn{1}{c|}{1-D in $\R^4$} & Unstable   & measure 0          \\ \hline
	\end{tabular}
	\vspace{0.1in}
	\label{tab:equil}
\end{table}
%

Furthermore, we ran simulations to illustrate some additional properties for the 1-D problem. 
We fixed the initial condition of agent 1 at $x_1(0) = -10$, its initial speed at $v_{01}=2$, and 
swept the initial condition and speed of agent 2 in 0.01 unit increments over the range $x_2(0) \in [-11, -8]$
and $v_{20}\in [1,3]$. Where applicable, the disturbance states $w_i$ are initialized at 0. As a measure of performance, 
we consider the extra time the agents took to clear the intersection. For one agent it is defined as the times 
it took to clear the intersection in the specific simulation run
minus what it would have taken with no interference from the other agent. 
If the agent did not clear in $20$ seconds, that was the time used:
\[ t_i^{ext} = \max\{20, {\rm arg min}_t\{x_i(t)\ge 0\} \}  -x_i(0)/v_{i0}, \ i=1,2 \]
The extra time plotted in the figures below is the sum of the two: $t^{ext} = t^{ext}_1 + t^{ext}_2$. 

\begin{figure}[htbp!]
\vspace{-.1in}
    \centering
\includegraphics[scale=0.48]{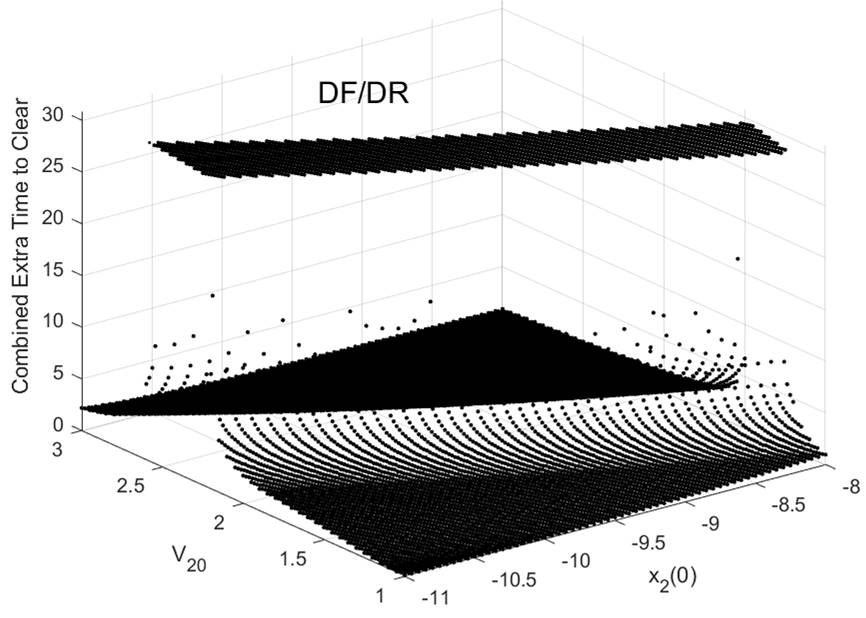}
    \vspace{-0.1in}
    \caption{Decentralized Reciprocal policy: the extra time it takes for both agents to clear the intersection due to mutual interference.}
    \vspace{-0.05in}
    \label{fig:dr_3d}
\end{figure}

To avoid the potential singularity at $x_i = 0$ for the DR algorithm, we have added the slack variable to the QP (\ref{dr_a1})
producing the dynamics 
\beq \barr{l}{ \dot x_1 = - \frac{v_{10}/M - x_1 \lambda h}{1/M + 4x_1^2}  \\*[2mm]
  \dot x_2 = - \frac{v_{20}/M - x_2 \lambda h}{1/M + 4x_2^2}   }  \label{dr_slack} \eeq
  When the weight $M$ on the slack variable goes to infinity, the original dynamics (\ref{dr_dynamic}) is recovered. For the DF/DR simulations shown in Figure \ref{fig:dr_3d},
  we have used $M = 10^6$. 

\begin{figure}[htbp!]
    \centering
\includegraphics[scale=0.46,trim=0 0 5 0, clip]{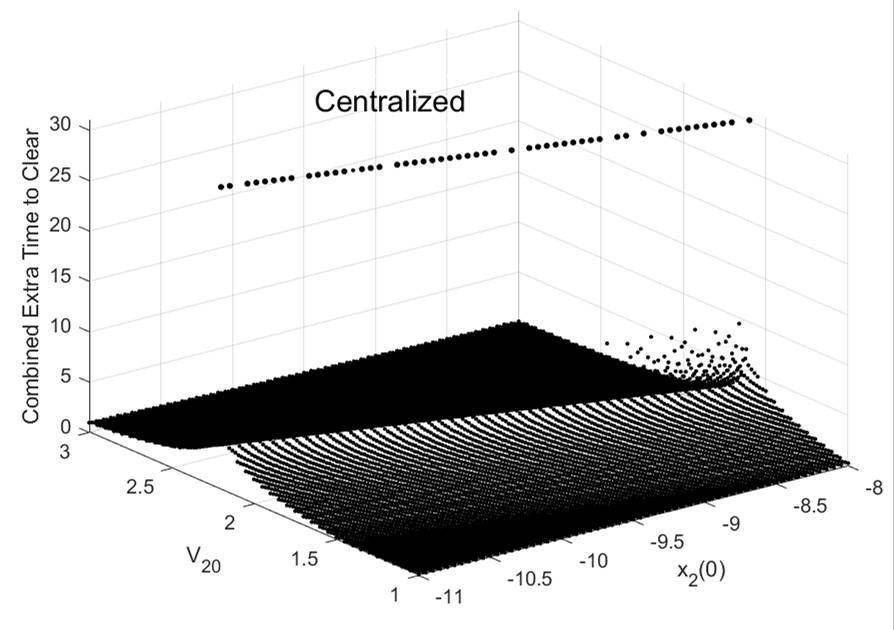}
    \vspace{-0.2in}
    \caption{Centralized policy: the extra time it takes for both agents to clear the intersection due to mutual interference.}
    \vspace{-0.05in}
    \label{fig:central_3d}
\end{figure}

The same initial condition and velocity sweep conducted with the Centralized policy produced the results shown in 
Figure \ref{fig:central_3d}. One can clearly see the combinations that lead to the equilibrium, i.e. those that start on the stable manifold corresponding to the 
selected $v_{10}$ and $v_{20}$.

\begin{figure}[htbp!]
\vspace{-.1in}
    \centering
\includegraphics[scale=0.46, trim = 0 0 2 5, clip]{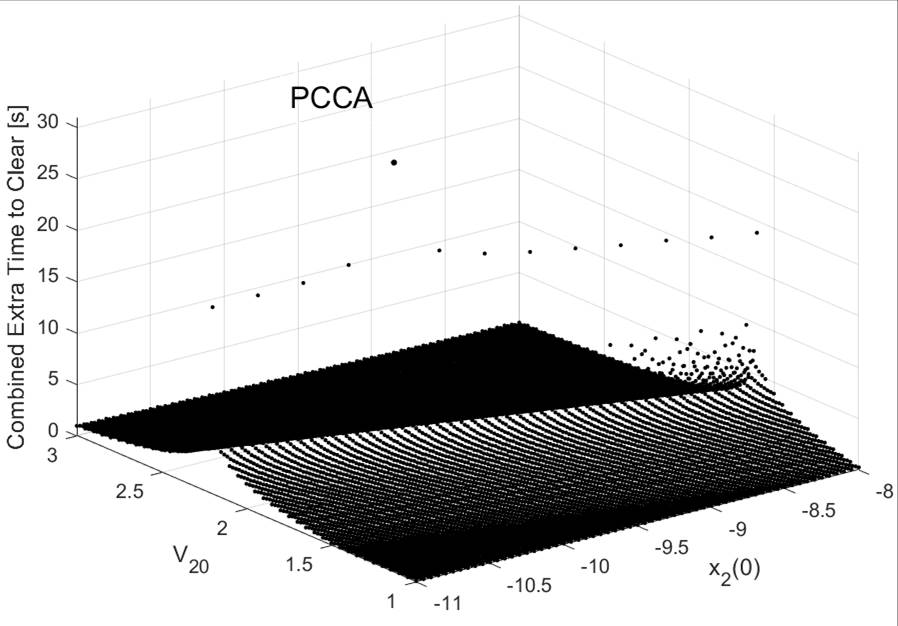}
    \vspace{-0.2in}
    \caption{PCCA policy: the extra time it takes for both agents to clear the intersection due to mutual interference.}
    \vspace{-0.05in}
    \label{fig:pcca_3d}
\end{figure}

The PCCA results are shown in Figure \ref{fig:pcca_3d}. It was somewhat of a surprise to observe  that there are far fewer PCCA than Centralized runs
that gridlock: 0.002\% PCCA and 0.1\% Centralized (for reference, DR policy gridlocks, i.e. reaches the max time of 20 seconds without either agent clearing the intersection, in 15.4\% of runs). One likely explanation is that, with fixed velocities, the stable manifold for the 
Centralized policy is a 1D line in the 2D plane, while for the PCCA  it is a 2D manifold in the 4D space. Obviously, it is less likely to hit the stable
manifold for the latter. Indeed, it happened only once in 60,000 points with completely symmetric initial conditions and nominal speeds. 

Finally, the CCS simulations have not been run because they would have always gridlocked, i.e. reached the 20s limit. 

\section{Conclusions}
This paper compared several CBF-based control policies for their performance in multi-agent scenarios. The results show that algorithms that take all the constraints into account and have or pretend to have control over all the agents, have lower convergence times and, as proven in this paper, are always feasible. The Centralized and PCCA policies showed minimal violations while the Decentralized Follower and Reciprocal methods had a few larger violations due to infeasibility. The CCS algorithm, structurally close to PCCA, showed liveness behavior more similar to Decentralized policies -- lower mean convergence time, but about the same number of gridlocks.  To explain the observed behavior, we analyzed the policies applied to a simpler 2-agent problem. It turned out that the policies exhibiting good liveness and lack of gridlock in simulation have unstable equilibria in the joint agent space, 
while policies with observed gridlock have stable equilibria. Beyond the retroactive analysis provided in this paper, it is not known to the authors how to design a control policy that creates unstable equilibria or modify an existing one to have this property. 
\noindent


\end{document}